\documentclass[twocolumn]{aastex62}
\usepackage{natbib}
\usepackage{apjfonts}
\usepackage{amsmath}
\usepackage{appendix}
\usepackage{todonotes}
\usepackage{outlines}

\DeclareSymbolFont{cmletters}{OML}{cmm}{m}{it}
\DeclareMathSymbol{v}{\mathalpha}{cmletters}{"76}

\newcommand{\aeq}{\ensuremath{a_{\rm eq}}}
\newcommand{\aeqval}{\ensuremath{0.13}}
\newcommand{\msun}{\ensuremath{M_{\odot}}}
\newcommand{\ein}{\ensuremath{e_{\rm in}}}
\newcommand{\lin}{\ensuremath{l_{\rm in}}}
\newcommand{\rg}{\ensuremath{r_{\rm g}}}
\newcommand{\rhor}{\ensuremath{r_{\rm H}}}
\newcommand{\mbh}{\ensuremath{M_{\rm BH}}} 

\newcommand{\mdot}{\ensuremath{\dot{M}}}
\newcommand{\mdotRM}{\ensuremath{\dot{m}}}

\graphicspath{{./}{figures/}}

\received{\today}
\revised{xxx}
\accepted{xxx}
\shorttitle{BH spin evolution in neutrino-cooled collapsars}
\shortauthors{Issa et al.}

\begin{document}

\title{Collapsar Black Hole Spin Evolution in 3D Neutrino Transport GRMHD Simulations}

\correspondingauthor{Danat Issa}
\email{danat@u.northwestern.edu}

\author[0009-0005-2478-7631]{Danat Issa}
\affiliation{Center for Interdisciplinary Exploration \& Research in Astrophysics (CIERA), Physics and Astronomy, Northwestern University, Evanston, IL 60201, USA}

\author[0000-0002-2875-4934]{Beverly Lowell}
\affil{Center for Interdisciplinary Exploration \& Research in Astrophysics (CIERA), Physics and Astronomy, Northwestern University, Evanston, IL 60201, USA}

\author[0000-0003-2982-0005]{Jonatan Jacquemin-Ide}
\affil{JILA, University of Colorado and National Institute of Standards and Technology, 440 UCB, Boulder, CO 80309-0440, USA}
\affil{Center for Interdisciplinary Exploration \& Research in Astrophysics (CIERA), Physics and Astronomy, Northwestern University, Evanston, IL 60201, USA}

\author[0000-0003-4475-9345]{Matthew Liska}
\affil{Center for Relativistic Astrophysics, Georgia Institute of Technology, Howey Physics Bldg, 837 State St NW, Atlanta, GA, 30332, USA}
    
\author[0000-0002-9182-2047]{Alexander Tchekhovskoy}
\affil{Center for Interdisciplinary Exploration \& Research in Astrophysics (CIERA), Physics and Astronomy, Northwestern University, Evanston, IL 60201, USA}
\affil{NSF-Simons AI Institute for the Sky (SkAI), 172 E. Chestnut St., Chicago, IL 60611, USA}

\begin{abstract}
Collapsars -- massive stars whose cores promptly collapse into black holes (BHs) -- can power long-duration gamma-ray bursts (LGRBs) via relativistic, collimated, electromagnetically-driven outflows, or jets. Their power depends on the BH magnetic field strength and spin.
To survive the infalling stellar material, jets need the central BH to attain dynamically important magnetic fields that can suppress the mass inflow and lead to a magnetically arrested disk (MAD). Previous work found that non-radiative MADs can spin down their BHs to an equilibrium spin, $\aeq^\text{nr}=0.035-0.07$. Such low spins result in extremely low power jets that may struggle to escape out of the star. However, the dense and hot collapsar disks emit neutrinos that cool the disk, reduce its thickness, and increase the angular momentum supply to the BH. 
Using 3D two-moment neutrino-transport general relativistic magnetohydrodynamic simulations, we show for the first time that successful collapsar jets powered by neutrino-cooled disks still rapidly spin down their BHs, although to a higher $\aeq\approx\aeqval$. This value is consistent with LIGO/Virgo/KAGRA inferred spins, is $2-4$x higher than for non-radiative MADs, and results in $4-16$x more powerful LGRB jets, which are more capable of drilling out of the progenitor star. This value of $\aeq$ holds across a wide range of progenitor structures and mass accretion rates, 
$\mdotRM\sim(0.1-10)\msun/\rm{s}$. We find that for typical LGRB durations, $t\gtrsim30$~s, such BHs consume sufficient mass to reach $\aeq\approx\aeqval$ by LGRB's end. However, shorter or lower-$\mdotRM$ LGRBs can leave behind more rapidly spinning BHs. 
\end{abstract}


\section{Introduction} \label{sec:intro}

Long-duration gamma-ray bursts (LGRBs) are highly energetic astrophysical events observed in distant galaxies and associated with the death of massive stars. They can be powered by collapsars \citep{Woosley_1993ApJ, Paczynski1998ApJ...collapsar}, in which a massive star undergoes gravitational collapse and forms a black hole (BH). If the stellar mantle possesses sufficient angular momentum, a magnetized accretion disk forms around the BH, and can launch a pair of relativistic jets, whose emission powers GRBs. 
The GRB jets are likely powered electromagnetically, via the Blandford-Znajek mechanism \citep[BZ,][]{BZ_1977}. In the latter case the jets are powered by the extraction of BH spin energy. The BZ jet power, an estimate of the observed GRB luminosity, is given by \( L_{\rm BZ} \propto a^2 \Phi^2 \), where \( a \) is the BH spin and \( \Phi \) is the magnetic flux at the BH horizon. Thus, these engine properties should dictate the GRB luminosity.

The BH spin distribution has been recently probed through gravitational waves (GW) from binary BH (BBH) mergers detected by LIGO/VIRGO/KAGRA. Most studies find spin distributions that are centered around low values (\( a \leq 0.3 \)), with highly spinning BHs (\( a \geq 0.6 \)) being rare or nonexistent \citep{the_ligo_scientific_collaboration_gwtc-3_2021,galaudage_building_2021,the_ligo_scientific_collaboration_population_2022,callister_no_2022,tong_population_2022,edelman_cover_2023}. Collapsar BHs, which power GRBs, are believed to account for a significant fraction (\(\sim 10\%-20\%\)) of BBH merger progenitors \citep{bavera_probing_2022,wu_are_2024}. Thus, GW observatories might offer stringent constraints on the final BH spin at the end of the GRB event. However, directly measuring the magnetic flux at the BH event horizon appears impossible, at least observationally.

Using global general relativistic magnetohydrodynamic (GRMHD) simulations, \cite{gottlieb_black_2022} found that to launch the jets in the face of the onslaught by the collapsing envelope, the BH needs to have strong, dynamically important magnetic fields, i.e., the central engine needs to be in or close to a magnetically arrested disk \citep[MAD,][]{Tchekhovskoy_MAD_2011MNRAS.418L..79T} state, consistent with analytical arguments \citep{Komissarov_2009MNRAS.397.1153K}. MADs are the natural steady state of accretion disks with abundant magnetic flux. As the flux is advected inward, it accumulates on the BH event horizon, reaching its saturation limit \citep{Tchekhovskoy_MAD_2011MNRAS.418L..79T,tchekhovskoy_magnetic_2015,jacquemin-ide_magnetic_2021,jacquemin-ide_magnetorotational_2024}. MADs produce the jet of power,
\begin{equation}
    P_{\rm jet} \sim a^2 \mdotRM c^2 \sim 10^{50}\, \text{erg/s} \left( \frac{a}{0.1} \right)^2 \left( \frac{\mdotRM}{0.01 \msun/ \text{s}} \right),
\end{equation}
where $a$ is the dimensionless BH spin parameter, $\mdotRM$ is the BH mass accretion rate, and $10^{50}\, \rm{erg/ s}$ is a typical LGRB luminosity \citep{goldstein_estimating_2016}; with the main uncertainty being the unknown gamma-ray radiative efficiency \citep{Nemmen_efficiency_jets_2012Sci...338.1445N}. This implies that typical jet powers require very small BH spin values, $a\sim 0.1$. Such low spins may seem surprising, especially given that GRB central engines are thought to occupy extreme and rare regions of the parameter space, often thought to be characterized by high spin and magnetic field strength values. Supporting this conclusion, \cite{Gottlieb2023ApJ...natalspin} used global GRMHD simulations to show that only collapsars with initially low-spin BHs (\( a \lesssim 0.2 \)) can reproduce the observed jet power distribution. Notably, these low spins align with the spin distributions measured by GW observatories, as discussed above.

Even if a collapsar BH begins with a maximal spin (\( a \sim 1 \)), dynamically important magnetic fields could rapidly spin it down to low values, with \(\aeq \sim 0.07\) in the highly super-Eddington limit \citep{Lowell2024_spindown} and \(\aeq \sim 0.035\) in the highly sub-Eddington regime \citep{narayan_jets_2022}. Here, $\aeq$ represents the equilibrium spin: once the spin reaches $\aeq$, it no longer changes. 
This rapid spin evolution can significantly impact jets' power and their activity during the stellar collapse and throughout the GRB duration.

\cite{Jacquemin-Ide2024ApJ...collapsarspindown} applied the spin evolution model of \cite{Lowell2024_spindown} to study how it impacts the jet power and jet breakout time during a GRB. They found that at typical accretion rates ($\mdotRM \sim 10^{-2}\,\msun/\rm{s}$), BH spins are expected to settle at the equilibrium value, \(\aeq \sim 0.07\), often before jet breakout. Indeed, we find that BH spin-down can happen within $t\sim 5\,\rm{s}$ for an accretion rate $\mdotRM \sim M_{\odot}/\rm{s}$.
This result supports the idea that collapsar-born BHs tend to have low spins, as even initially rapidly-spinning BHs spin down. Additionally, they showed that an initially rapidly-spinning BH would produce jets that are too powerful compared to observations, which limits the initial spin to \( a \leq 0.4 \).

Building on the models of \cite{Lowell2024_spindown} and \cite{Jacquemin-Ide2024ApJ...collapsarspindown}, \cite{Wu_2024_GRB_spinevol} argue that GRBs are unlikely to be powered by a MAD. Instead, they suggest that jets of higher efficiency could be launched at lower, sub-MAD level of BH saturation with magnetic flux, because of the higher equilibrium spin (\(a \sim 0.5\)) at that flux. Moreover, they propose that most LGRBs originate at even lower BH saturation levels, which may contradict the idea that a near-MAD state is required for jet launching \citep{Komissarov_2009MNRAS.397.1153K,gottlieb_black_2022}.

So far, collapsar BH spin evolution has been simulated for non-radiative adiabatic thick disks. However, due to large mass supply in massive collapsing stars, these disks are expected to be cooled by neutrinos \citep{Chevalier_1989ApJ,Batta_2014MNRAS,Siegel_2019Natur}. Previously, neutrino cooling has not been included in collapsar simulations that study BH spin evolution, as it requires computationally expensive neutrino transport. Moreover, in collapsars, neutrinos are crucial for modeling the conditions for heavy-element nucleosynthesis \citep[see][for a review]{SiegelReview2022}. This further underscores the importance of modeling their effects. Neutrino cooling can alter disk thermodynamics and structure, reducing disk thickness \citep{chen_beloborodov_2007}, which, in turn, affects angular momentum transport and BH spin evolution.  

This effect has been observed in radiative and cooled MAD simulations at sub- and near-Eddington rates, 
where the disk thickness strongly influences both spin-down efficiency and equilibrium spin. \cite{ricarte_recipes_2023} found that as the disk cools, equilibrium spin increases significantly, reaching \(\aeq \sim 0.8\) for BHs accreting at the Eddington rate. Similarly, \cite{Lowell_prep} (in prep) reported that spin-down becomes less effective for thinner disks, though to a lesser extent than in \cite{ricarte_recipes_2023}, finding \(\aeq \sim 0.3\) for \(h/r = 0.1\).

This motivates our goal to shed light on how neutrino cooling affects the evolution of BH spin. We present the first suite of 3D GRMHD simulations of collapsars with neutrino transport to study collapsar BH spin evolution. We introduce the BH spin evolution model in Section \ref{sec:spin_evol}. In Section \ref{sec:num_setup} we describe our numerical setup. In Section \ref{sec:results} we present the simulation results and compute fits for the parameters of the spin evolution model. Using this model, we make predictions for spin evolution on BH engine activity timescales in Section \ref{sec:spin_evol_analysis}. Finally, in Section \ref{sec:discussion}, we discuss the broader implications of the BH spin evolution on the GRB production and observations. In this paper, we adopt the units $G=c=1$, and use Heaviside-Lorentz units for magnetic fields and thereby absorb the factor of $1/\sqrt{4\pi}$ into the definition of magnetic field strength, $B$.

\begin{table*}[!htbp]
    \centering
    \caption{The summary of the simulation parameters. Leftmost column shows the model name, which follows [pc]\#[N0] naming convention - first letter stands for the initial density profile, where "p" stands for power-law density profile and "c" - constant-density core; "\#" - denotes BH spin ($10\times a$); "N0" - denotes models with no neutrino transport. Following columns, from left to right: BH spin, neutrino transport, 3D grid resolution, final time in the units of $\rg/c$ and seconds, time-averaging window for the parameters in the MAD spin evolution model, specific angular momentum, $l_{\rm MAD}$, energy fluxes, $e_{\rm MAD}$, in the MAD state averaged in $t,\theta,\varphi$, measured on the horizon, and the MAD spin-up parameter, $s$.}
    \begin{tabular}{|r@{}l|c|c|c|c|c|c|c|c|c|}
    \hline
    Mo&del & $a$ & $\nu$-transport & $N_{\rm r} \times N_\theta \times N_\varphi$ & $t_{\rm f}$ $[10^4 \rg/c]$ & $t_{\rm f}$ [s] & $t_{\rm avg}$ $[10^4 \rg/c]$ & $l_{\rm MAD}$ & $e_{\rm MAD}$ & $s$\\
    \hline
        c0& & 0.0 & yes & $288 \times 288 \times 64$ & 4.96 & 0.98 & (3.00, 4.96) & 0.985 & 0.773 & 1.045 \\
        c1& & 0.1 & yes & $288 \times 288 \times 64$ & 4.91 & 0.97 & (3.00, 4.91) & 0.941 & 0.776 & 0.097 \\
        c1&N0 & 0.1 & no & $288 \times 288 \times 64$ & 7.87 & 1.56 & (3.00, 7.87) & 1.252 & 0.982 & 0.296 \\
        c2& & 0.2 & yes & $288 \times 288 \times 64$ & 5.04 & 1.00 & (3.00, 5.04) & 0.811 & 0.761 & -0.877 \\
        c5& & 0.5 & yes & $288 \times 288 \times 64$ & 5.92 & 1.17 & (3.00, 5.92) & 0.785 & 0.736 & -3.174 \\
        c8& & 0.8 & yes & $288 \times 288 \times 64$ & 5.89 & 1.16 & (3.00, 5.89) & 0.494 & 0.674 & -5.179 \\
        p1& & 0.1 & yes & $192 \times 128 \times 64$ & 13.70 & 2.71 & (3.00, 13.70) & 1.474 & 0.904 & 0.675 \\
        p1&N0 & 0.1 & no & $192 \times 128 \times 64$ & 12.71 & 2.51 & (3.00, 12.71) & 1.435 & 0.979 & 1.014 \\
        p2& & 0.2 & yes & $192 \times 128 \times 64$ & 10.36 & 2.05 & (3.00, 10.36) & 1.228 & 0.880 & -0.933 \\
        p5& & 0.5 & yes & $192 \times 128 \times 64$ & 8.35 & 1.65 & (3.00, 8.35) & 0.981 & 0.865 & -4.320 \\
        p8& & 0.8 & yes & $192 \times 128 \times 64$ & 4.95 & 0.98 & (3.00, 4.95) & 0.547 & 0.809 & -7.939 \\
    \hline
    \end{tabular}
    \label{tab:models}
\end{table*}

\section{BH Spin Evolution Model}
\label{sec:spin_evol}

We define the dimensionless BH spin, $a$, which ranges from $-1$ for maximally spinning retrograde BHs to $+1$ for prograde BHs: 
\begin{equation} \label{eq:spin_def}
    a = \frac{J}{M^2},
\end{equation}
where $J$ and $M$ are the angular momentum and the mass-energy of the BH, respectively. As the BH accretes, $J$ and $M$ evolve by accreting angular momentum and energy such that
\begin{equation}
    \dot{J} = \mdotRM M \lin, \quad \dot{M} = \mdotRM \ein,
\end{equation}
where $\lin, \ein$ are specific angular momentum and energy fluxes, $\mdotRM$ is the rest mass accretion rate (not to be confused with $\mdot$ - the total mass-energy accreted), all evaluated at the BH event horizon. Here, dots represent time derivatives. We can produce the equations for time evolution of the spin by taking the time derivative of $a$ \citep[e.g.,][]{moderski_black_1996,Lowell2024_spindown}:
\begin{equation}
    \begin{split}
    \frac{da}{dt} &= s(a, \lin, \ein) \frac{\mdotRM}{M} \\
    \frac{dM}{dt} &= \ein \mdotRM
    \end{split}
\end{equation}
where $s$ is the dimensionless ``spin-up'' parameter \citep{Gammie2004},
\begin{equation}
    s = l_{\rm in} - 2 a e_{\rm in}.
    \label{eq:spinup_parameter}
\end{equation}
While the accreted angular momentum increases $s$, and spins up the BH, the accreted energy spins down the BH. As the BH accretes mass and angular momentum, its spin approaches the equilibrium value. Once the BH accretes enough mass, the spin-up parameter vanishes, and the BH reaches the equilibrium spin.

In the simulations, we measure the BH mass accretion rate, $\mdotRM$, and specific fluxes $\ein, \lin$, as
\begin{equation}
    \mdotRM (r) = -\int \rho u^r dA_{\theta \varphi}, 
    \label{eq:fM}
\end{equation}
\begin{equation}
    \lin(r) = \frac{1}{M} \frac{\dot{J}(r)}{\mdotRM (r)} = \frac{1}{M} \frac{1}{\mdotRM (r)} \int T^r_\varphi dA_{\theta \varphi},
    \label{eq:fLoverfM}
\end{equation}
and 
\begin{equation}
    \ein(r) = \frac{\dot{M}(r)}{\mdotRM (r)} = \frac{1}{\mdotRM (r)} \int T^r_t dA_{\theta \varphi},
    \label{eq:fEoverfM}
\end{equation} 
where we take the integrals over surfaces of constant $r$, where $dA_{\theta\varphi} = \sqrt{-g} d\theta d\varphi$ is the surface area element, and $r$, $\theta$, $\varphi$ are radius, polar and azimuthal angles in Kerr-Schild coordinates. Here, $T^\mu_\nu$ is the stress-energy tensor,
\begin{equation}
    T^\mu_\nu = (\rho + u_{\rm g} + p_{\rm g} + b^2) u^\mu u_\nu + \left(p_{\rm g} + \frac{b^2}{2} \right) \delta^\mu_\nu - b^\mu b_\nu + R^\mu_\nu,
    \label{eq:total_EMtensor}
\end{equation}
and $\rho$, $u_{\rm g}$, $p_{\rm g}$, $u^\mu$ are gas mass density, internal energy, pressure, and contravariant 4-velocity vector; $\delta^\mu_\nu$ is Kronecker delta; $b^\mu$ is contravariant magnetic field 4-vector, such that $b^2=b^\mu b_\mu$. In Eq.(\ref{eq:total_EMtensor}), $R^\mu_\nu$ is the total neutrino radiation energy-momentum tensor,
\begin{equation}
    R^\mu_\nu = \sum_{\rm s\,=\,\nu_{\rm e}, \bar{\nu}_{\rm e}, \nu_{\rm x}}\frac{1}{3} E_{\rm rad, (s)} \left( u_{\rm rad, (s)}^\mu u_{\rm rad, (s), \nu} + \delta^\mu_\nu \right),
    \label{eq:nu_EMtensor}
\end{equation}
where $E_{\rm rad}$ is the neutrino energy density, measured in the reference frame, whose 4-velocity is $u_{\rm rad}^\mu$, where radiation is isotropic. The sum is taken over all neutrino species evolved in simulation: electron neutrinos, antineutrinos and heavy lepton neutrinos.

We compute $\mdotRM$, $\lin$, and $\ein$ at the event horizon, $r=\rhor = 1+(1-a^2)^{1/2}$, where careful consideration is required due to numerical floors. GRMHD codes cannot handle vacuum, therefore we must set a ceiling on the value of magnetization, $\sigma=b^2/\rho$, or, equivalently, a floor of $\rho$, which results in nonphysical density and internal energy addition near the BH event horizon. In our simulations, we limit the magnetization to not exceed $\sigma_{\max} = 25$. Similar to \cite{Lowell2024_spindown}, we find that using the cutoff on magnetization, $\sigma < 2 \sigma_{\max} /3$, in the integrals in Eq.(\ref{eq:fM},\ref{eq:fLoverfM},\ref{eq:fEoverfM}), allows us to remove the contribution of the floors to these fluxes near the BH.

\section{Numerical Setup} 
\label{sec:num_setup}

We performed a suite of simulations using the graphical processing unit (GPU)-accelerated neutrino-GRMHD code $\nu$H-AMR \citep{Issa2024arXiv241002852I...nuHAMR}. The numerical implementation of ideal GRMHD module is described in \cite{Liska2022ApJS..263...26L...hamr}. Neutrino transport implementation follows energy-integrated ("gray") M1 scheme, where 3 species of neutrinos -- electron neutrinos, $\nu_{\rm e}$, electron antineutrinos, $\bar{\nu}_{\rm e}$, and all of the heavy lepton neutrinos combined, $\nu_{\rm x}$, -- are treated in fluid-like manner. For the details of the numerical implementation of the neutrino transport scheme, see \cite{Issa2024arXiv241002852I...nuHAMR}. For an equation of state (EOS), the code uses Helmholtz EOS \citep{timmes_eos_2000ApJS..126..501T}. 

In this section, we describe the initial setup of the collapsar simulations, which follows \cite{Issa2024arXiv241002852I...nuHAMR}. Table~\ref{tab:models} summarizes the key simulation parameters. Initially, the spatial distribution of the stellar gas is spherically symmetric and depends only on the spherical polar radius $r$:
\begin{equation}
    \rho(r) = \rho_0 r^{-\alpha_{\rm p}} \left( 1 - \frac{r}{R_{\rm star}} \right)^{3},
\end{equation}
where $R_{\rm star} = 4\times 10^{10} \rm{cm}$ is the stellar radius and $\rho_0$ is the density normalization factor, which is chosen such that the total mass of the star (without the BH, whose mass is $M_{\rm BH}=4\msun$) is $M_{\rm star} = 70\,\msun$. To study the dependence of the BH spin evolution on the stellar structure of the collapsar progenitor, we employ two different values of the density slope, $\alpha_{\rm p}$. In models denoted by p\#, we set $\alpha_{\rm p} = 1.5$, motivated by the typical post-collapse density profiles (\cite{Halevi2023ApJ...corecollapse}), which result in mass accretion rates, $\mdotRM \sim 0.1-1\msun / \rm{s}$. We also consider a separate progenitor model, denoted by c\#, with the density profile that is more concentrated near the BH: constant density core of radius $R_{\rm core} = 10^8 \rm{cm}$ (where $\alpha_{\rm p}=0$), followed by a steeper density profile with $\alpha_{\rm p} = 2.5$ at $r>R_{\rm core}$. This density profile results in much higher mass accretion rates at early times, $\mdotRM \gtrsim{} 10 \msun / \rm{s}$.

We set the initial thermal pressure profile in the star to be:
\begin{equation}
	P_{\rm gas}(r) = P_{0} \frac{\rho(r)}{r / \rg} ,
\end{equation}
where $P_{0} = 0.1$ and $P_{0} = 0.25$ for models with and without the constant density core, respectively. The initial thermal pressure does not impact the disk dynamics since the pressure in the disk will be dominated by energy dissipation due to turbulence associated with the accretion process, and the chosen prefactors ensure that the pressure on the grid is within the temperature validity range of the Helmholtz EOS.

Initially the gas velocity is purely azimuthal. We choose the rotation profile such that each spherical shell has the same constant angular frequency whose radial profile we express via the specific angular momentum:
\begin{equation}
	l(r, \theta) = 
		\min \left[ 1, \left( \displaystyle \frac{r}{r_{\rm rot}} \right)^2 \right] \sqrt{\mbh r_{\rm circ}} \sin^2\theta ,
\end{equation}
where $r_{\rm circ} = 25 \rg$ is the circularization radius. Here, we choose the radius of rigid rotation, $r_{\rm rot} = 70\rg$, such that the flow is sub-Keplerian inside $r_{\rm rot}$. This facilitates disk formation once the shell at this radius reaches the BH horizon on the free-fall timescale $t_{\rm f-f} \approx r_{\rm rot}^{1.5}/\sqrt{2}\approx 8 \rm{ms}$. This is consistent with the expectations for collapsar progenitors \citep{Gottlieb2024ApJ...976L..13G}. The equatorial inflow faces a centrifugal barrier at $r = r_{\rm circ}$, whereas polar inflows fall onto the BH.

We choose the magnetic field strength and spatial distribution that are conducive for successful magnetized explosions, even at lower BH spins. We set the magnetic field through the covariant vector potential $A_{\varphi}$:
\begin{equation}\label{eq:Aphi}
	A_{\varphi} (r, \theta) = \mu \sin^2\theta \times \max \left[ \frac{r^2}{r^2 + R_{\rm core}^2} - \left( \frac{r}{R_{\rm star}} \right)^3, 0 \right]\,,
\end{equation}
where we adopt a stellar core radius, $R_{\rm core} = 10^8 \,\rm{cm}$. This configuration results in a nearly uniform vertical magnetic field at $r \lesssim R_{\rm core}$ that turns radial inside the star and closes near the stellar surface. The magnetic moment $\mu$ is chosen so that magnetization in the core is $\sigma={b^2}/{\rho}\lesssim 0.1$, corresponding to the gas to magnetic pressure ratio of $\beta = \max P_{\rm gas} / \max P_{\rm mag} \sim$ few.

We use a uniform computational grid in $\log_{10} r$, $\theta$, and $\varphi$ coordinates that spans $0.15\le\log_{10}(r/\rg)\le5, 0\le\theta\le\pi, 0\le\varphi\le2\pi$, with outflow boundary conditions in $r$, and transmissive boundary conditions in $\theta$, and periodic in $\varphi$. We provide the numerical resolution of our simulations in Table~\ref{tab:models}.

For each of the two types of stellar progenitors, we perform a suite of simulations at different fixed BH spins $a$, spanning the range of different possible spins of collapsar BHs. Additionally, to evalute the contribution of neutrino transport to the spin evolution, for each of the progenitors, we run simulations at $a=0.1$, where we switch off neutrino transport.

\section{Numerical results}
\label{sec:results}

\subsection{Accretion \& Outflows}

\begin{figure}[!tbp]
    \centering
    \includegraphics[width=\columnwidth]{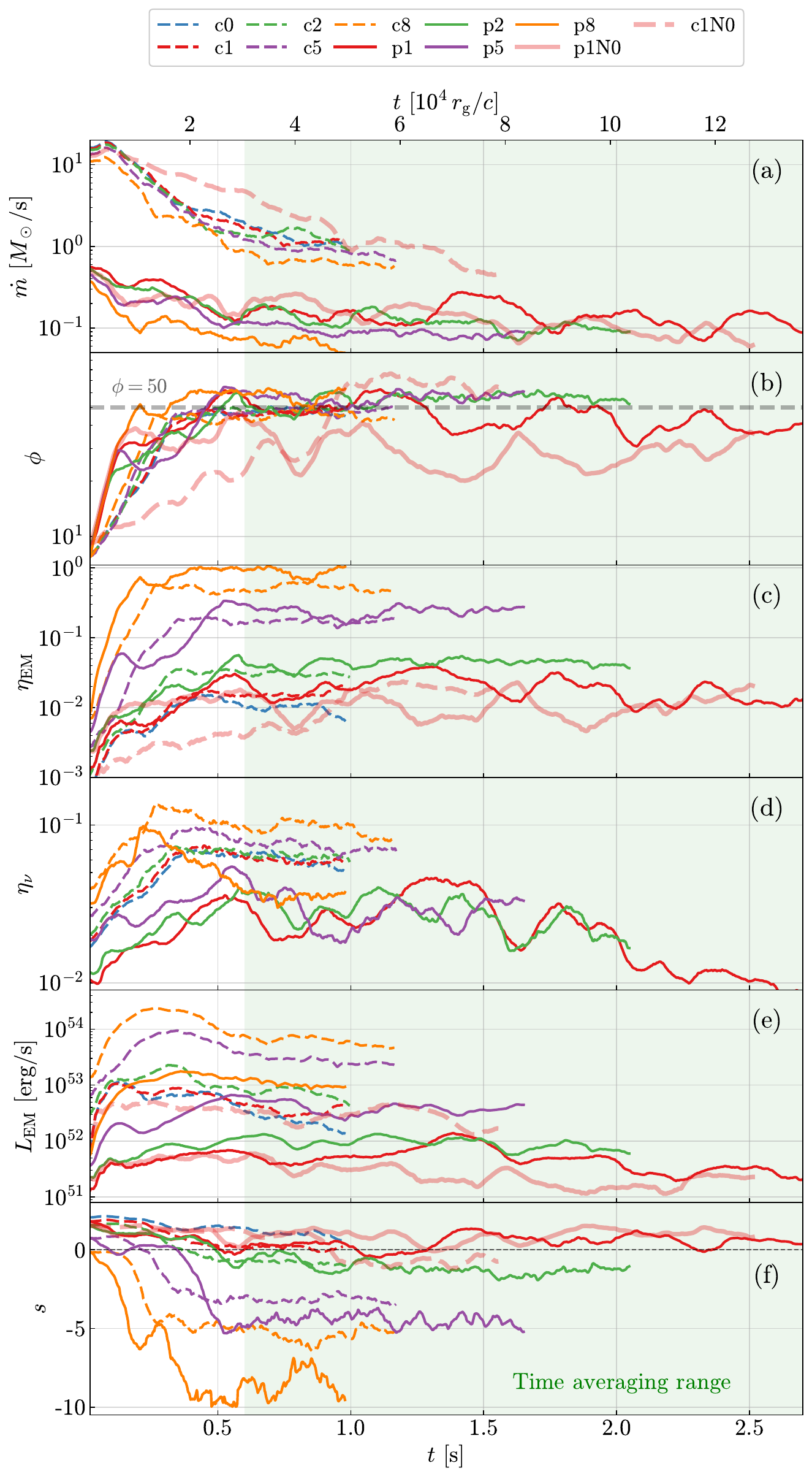}
    \caption{We consider two different collapsar progenitor structures for a range of BH spin. Progenitors with constant density cores (models c\#, where \# is BH spin) are shown with dashed lines, and more typical progenitors  (models p\#) with power-law density profiles, $\rho \propto r^{-1.5}$, are shown with solid lines. 
    Mass accretion rates, $\mdotRM$, (\textbf{panel [a]}) reach $20 \msun / \rm{s}$ in c\# models, compared to $\sim 0.1-0.5\msun / \rm{s}$ in p\# models. 
    Dimensionless magnetic flux, $\phi$, (\textbf{panel [b]}) computed at the BH horizon, saturates at $\phi_{\rm MAD}\approx 50$, which is a typical value for a magnetically arrested disk (MAD). 
    Magnetically-driven outflow efficiencies, $\eta_{\rm EM}$ (\textbf{panel [c]}), peak after the BH saturates with magnetic flux and turn MAD, where the peak value is higher for higher spins (for the fit, see Eq.~\ref{eq:eta_jet_fit}. 
    Neutrino efficiencies, $\eta_{\rm \nu}$ (\textbf{panel [d]}), vary between $1-10\%$, depending on the spin and $\mdotRM$. 
    In models with rapidly spinning BHs and moderate-to-high $\mdotRM$, the power of electromagnetic outflows, $L_{\rm EM}$ (\textbf{panel [e]}), exceeds $10^{54}$ erg/s, which is much higher than observed values; this suggests that in nature collapsars that launch GRBs are likely slowly spinning. 
    The spin-up parameter, $s$ (\textbf{panel [f]}) defined in Eq.~\ref{eq:spinup_parameter}, becomes roughly constant in MAD state. We compute the average $s$ at $t\geq 0.6\,\rm{s}$ (marked in light green), when all models are in MAD state. For clarity, all quantities are averaged over a moving time window, $\tau=5\times 10^3 \rg/\rm{c}\approx 0.1\,\rm{s}$.}
    \label{fig:plt_vs_time}
\end{figure}

Figure~\ref{fig:plt_vs_time} shows the time-dependence of various quantities measured near the BH. We measure the mass accretion rate, $\mdotRM$ (Eq.~\ref{eq:fM}) at the BH event horizon, with corrections that account for numerical density and internal energy floors, described in Sec.~\ref{sec:spin_evol}. As we outlined in Sec.~\ref{sec:num_setup}, we consider two distinct collapsar progenitor structures, whose key difference lies in their initial density slopes. Fig.~\ref{fig:plt_vs_time}(a) shows that typical collapsar models, denoted by p\# (solid lines), result in mass accretion rates, $\mdotRM \sim 0.1-1 \msun / \rm{s}$, that gradually decrease in time. Models with initial constant density cores, denoted by c\# (dashed lines), reach much higher $\mdotRM \sim 10 \msun / \rm{s}$ at $t \sim 0.1 \rm{s}$, when the core gets accreted, after which $\mdotRM$ drops in time as a power law. There are factor of $\sim 2$ differences in the exact values of $\mdotRM$ due to differences in BH spins: more slowly spinning BHs accrete faster.

We compute the degree of BH saturation with magnetic flux, by measuring dimensionless magnetic flux, $\phi$, 
\begin{equation}
    \phi = \frac{ \Phi_{\rm BH} }{ \sqrt{ \langle\mdotRM\rangle_{\tau} } } = \frac{ 1/2 }{ \sqrt{ \langle\mdotRM\rangle_{\tau} } } \int_{r=\rhor} | B^{r} | dA_{\theta\varphi} ,
\end{equation}
where $\Phi_{\rm BH}$ is the absolute magnetic flux through one hemisphere, measured as half of the total absolute flux on the BH. We normalize it by $\sqrt{\langle\mdotRM\rangle_{\tau}}$, where $\langle\cdot\rangle_{\tau}$ denotes the rolling average over the time interval $\tau=5\times 10^3\, \rg/c$, $\langle\mdotRM\rangle_{\tau}$. Fig.~\ref{fig:plt_vs_time}(b) shows that $\phi$ saturates at the characteristic value, $\phi_{\rm MAD} \sim 50$: at this point, BH magnetic flux becomes dynamically-important, begins to obstruct the gas infall, and enters the magnetically arrested disk (MAD) state \citep{Tchekhovskoy_MAD_2011MNRAS.418L..79T}. In fact, MAD-level strength BH magnetic fields appear to be needed in collapsar powered-GRBs, for the magnetically-driven BH outflows to survive the onslaught of the infalling stellar material \citep{Komissarov_2009MNRAS.397.1153K,gottlieb_black_2022, Issa2024arXiv241002852I...nuHAMR}. 

The mass accretion rates set the energy budgets available for powering the transient, particularly, the jet power and neutrino luminosities, through their respective efficiencies: the EM outflows launching efficiencies, $\eta_{\rm EM}$, and neutrino emission efficiencies, $\eta_{\nu}$,
\begin{equation}
    \eta_{\rm x} = \frac{ L_{\rm x} }{ \langle\mdotRM\rangle_{\tau} }, \quad \text{where x} = \rm{EM}, \nu ,
    \label{eq:eta_jet}
\end{equation}
where electromagnetic (EM) outflow (or jet) power, $L_{\rm EM}$, is defined as,
\begin{equation}
    L_{\rm EM} = - \int T_{\text{EM}, t}^r dA_{\theta\varphi}  \Big|_{r=\rhor} = - \int \left( b^2 u^r u_t - b^r b_t \right) dA_{\theta\varphi}  \Big|_{r=\rhor} ,
\end{equation}
and neutrino luminosity, $L_{\rm \nu}$, measured at $r = 100\,\rg$, outside the main emission region, 
\begin{equation}
    L_{\rm \nu} = - \int R_{t}^r dA_{\theta\varphi} \Big|_{r=100\,\rg} ,
\end{equation}
where $R^{\mu}_{\nu}$ is defined in Eq.~\ref{eq:nu_EMtensor}.
Fig.~\ref{fig:plt_vs_time}(c) shows that $\eta_{\rm EM}$ depends on the degree of BH saturation with magnetic flux, $\phi$, and the BH spin, $a$,
\begin{equation}
    \eta_{\rm EM} (\phi, a) = \eta_{\phi}\eta_a \sim \left( \frac{\phi}{\phi_{\rm MAD}} \right)^2 \, \eta_a(a) ,
    \label{eq:eta_phi_a}
\end{equation}
such that $\eta_{\rm EM}$ plateaus once the BH is in the MAD state. Its maximum value depends on the BH spin $a$; we calculate the fits for $\eta_{a}(a)$ in Sec.~\ref{sec:fits} (Eq.~\ref{eq:eta_jet_fit}). 

\begin{figure}[hbtp]
    \centering
    \includegraphics[width=\columnwidth]{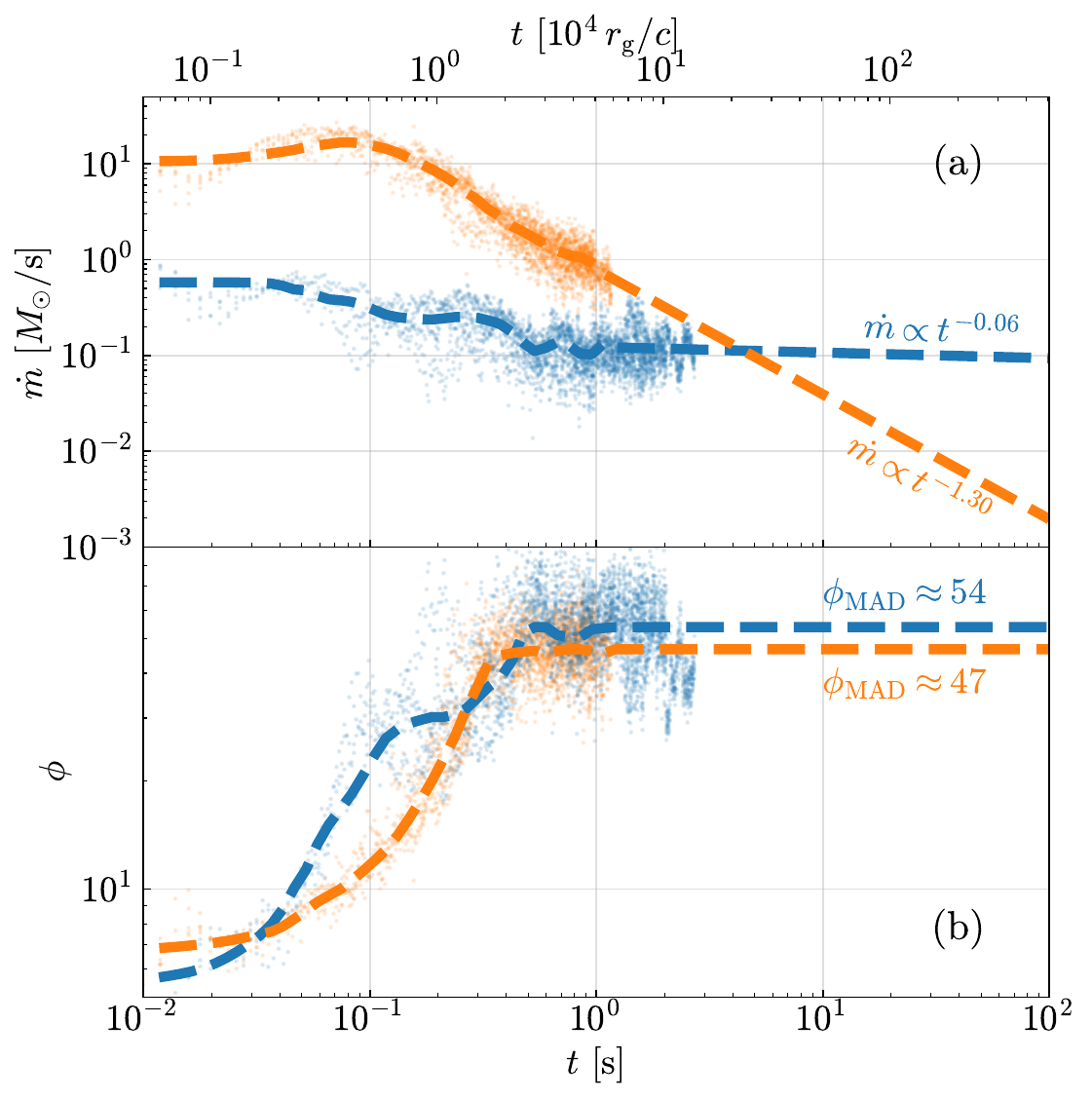}
    \caption{Mass accretion rates (\textbf{panel [a]}) and dimensionless fluxes (\textbf{panel [b]}) that begin with the same progenitor models, evolve similarly in time for a wide range of spins. Orange shows all c\# models (with constant density core), and blue - all p\# models (typical collapsars with $\rho (r) \propto r^{-1.5}$). We demonstrate profiles of $\mdotRM (t)$ and $\phi (t)$ for each model (faint markers), and compute the average profiles (thick dashed lines) for each of the two progenitor models. Beyond the simulation runtimes, $t \gtrsim{} t_{\rm sim} \sim 1\,\rm{s}$, $\mdotRM$ are extrapolated with the power-law fits in time,  $t^{-p}$, where the fits are calculated over $0.6\,\text{s} < t < t_{\rm sim}$, and $p=0.06$ and $1.30$ for p\# and c\# models respectively. We assume that the BHs remain saturated with magnetic flux, thus, we set $\phi$ to their MAD-levels, which are $\phi_{\rm MAD} \approx 54$ and $47$ for p\# and c\# models respectively. We use these extrapolated fits for the spin evolution analysis in Sec.~\ref{sec:spin_evol_analysis}.}
    \label{fig:mdot_phi_avg}
\end{figure}

Fig.~\ref{fig:plt_vs_time}(d) shows the total neutrino efficiencies $\eta_{\nu}$. The neutrino efficiencies lie within a range between $1-10\%$ and depend on the mass accretion rate. Higher mass accretion rates lead to more efficient neutrino cooling. Hence, it is unsurprising that c\# models (dashed lines) with higher $\mdotRM$ have higher $\eta_{\nu}$.

Fig.~\ref{fig:plt_vs_time}(e) shows jet power: it is the product of $\eta_{\rm EM}$ and $\mdotRM$ and exceeds $L_{\rm EM} \sim 10^{54} \rm erg/s$ for our high-$\mdotRM$ model c8 with BH spin $a=0.8$. Lower BH spin models launch jets with jet powers that are more consistent with GRB observations, $L_{\rm EM} \sim \text{few}\times 10^{51} \rm erg/s$. Once the magnetic flux saturates the BH, the jet power follows $\mdotRM$, as the efficiency, $\eta_{\rm EM}$, saturates at the maximum MAD value (Eq.~\ref{eq:eta_phi_a}).

Finally, Fig.~\ref{fig:plt_vs_time}(f) shows the spin-up parameter, $s$, defined in Sec.~\ref{sec:spin_evol} (Eq.~\ref{eq:spinup_parameter}), as a function of time. Before BH accumulates sufficient magnetic flux, the spin-up parameter is determined by the accretion of specific angular momentum, $\lin$, and energy, $\ein$. This leads to $s \geq 0$, in line with the standard thin disk spin evolution. In the absence of strong magnetically-driven outflows, these disks are known to spin up the BH to $\aeq \approx 1$ \citep{bardeen_kerr_1970,thorne_spin_1974ApJ...191..507T}. However, this picture significantly changes with the onset of MAD. As the flow approaches the MAD state, the spin-up parameter, $s$, decreases. This reflects the large-scale magnetic field on the BH exerting significant spin-down torques at spins $a \gtrsim{0.2}$, to drive the equilibrium spin down, qualitatively similar to \cite{Lowell2024_spindown}. Fig.~\ref{fig:plt_vs_time}(f) shows that even if $\mdotRM$ is not constant in our models, $s$ is approximately constant, since both $\lin,\,\ein$ are normalized by  $\mdotRM$ (Eq.~\ref{eq:fLoverfM},\ref{eq:fEoverfM}). 

To make sense of the landscape of our models, Fig.~\ref{fig:mdot_phi_avg} shows the scatter plot of all $\mdotRM(t)$ and $\phi(t)$ data points for c\# (orange dots) and p\# models (blue dots) $\nu$GRMHD models. The dashed orange and blue curves in Fig.~\ref{fig:mdot_phi_avg} show the averages of the data points for each of the c\# and p\# models, respectively. We calculate power law fits in time for $\mdotRM$, over the time window, $t > 0.6\,\text{s}$, and use them as extrapolation beyond the simulation durations. Over the same time period, we compute the average $\phi_{\rm MAD}$ in MAD state, which we extrapolate as a constant beyond the simulation runtimes. Here, we assume that as $\mdotRM$ decreases with time, the BHs remain in the MAD state throughout the duration of the BH engine activity \citep[e.g.,][]{Tchekhovskoy_2014MNRAS.437.2744T,Tchekhovskoy_2015MNRAS.447..327T,Fernandez_2019MNRAS.482.3373F,Christie_2019MNRAS.490.4811C}. For the $\mdotRM(t)$ extrapolation fits, we find that in p\# models, $\mdotRM \propto t^{-0.06}$, and much steeper $\mdotRM \propto t^{-1.3}$ in c\# models due to steeper density profiles. As for MAD saturation threshold, $\phi_{\rm MAD}$, p\# models plateau at a slightly different level, $\phi_{\rm MAD} \approx 54$, while c\# models stabilize at $\phi_{\rm MAD} \approx 47$. Both values are around the threshold of $\phi_{\rm MAD} \sim 50$, found in the simulations of thick MADs \citep[e.g.,][]{Tchekhovskoy_MAD_2011MNRAS.418L..79T, Lowell2024_spindown}. The time of the MAD onset, $t_{\rm MAD}$, is approximately the same across progenitor models, with c\# models achieving it earlier, at $t_{\rm MAD} \approx 0.3\,\rm{s}$, compared to $t_{\rm MAD} \approx 0.4\,\rm{s}$ for p\# models.

\subsection{Variability in lower spin models} 
\label{sec:wobbling}

\begin{figure*}[!tbp]
    \centering
    \includegraphics[width=0.7\textwidth]{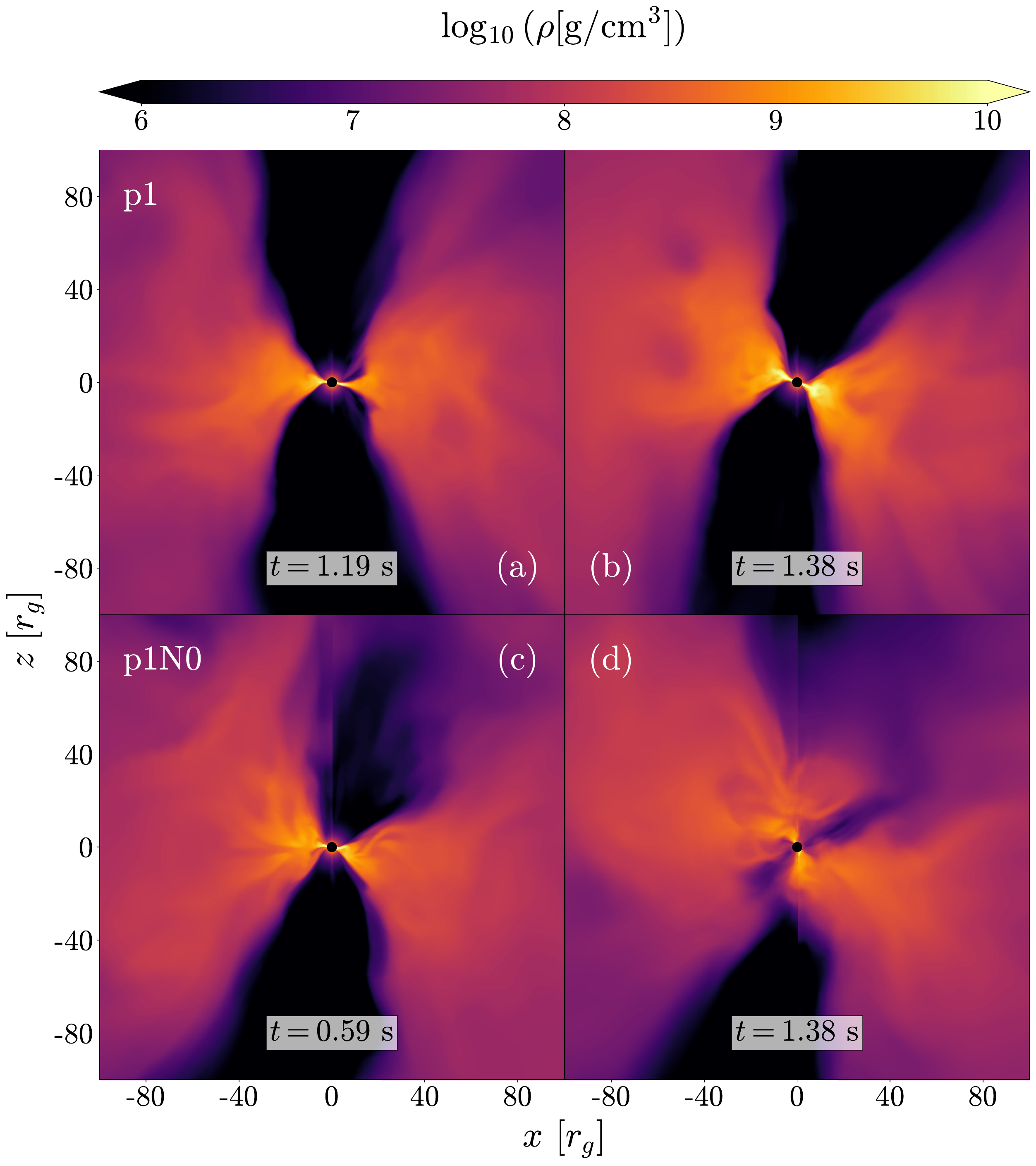}
    \caption{Collapsar disks around slowly spinning BHs develop strong wobbles with respect to the BH spin midplane, e.g., in models p1 (\textbf{panels [a-b]}) and p1N0 (\textbf{panels [c-d]}), both of which have $a = 0.1$. Here, we show meridional slices through the logarithm of gas density, at two different times. \textbf{panels [a,c]:} Before the onset of disk misalignment, the jets clear out the polar funnel along the BH spin rotation axis (vertical in the figure). Here, BH is saturated with the magnetic flux at the MAD level (see Fig.~\ref{fig:plt_vs_time}b). \textbf{panels [b,d]:} Jets powered by such slowly spinning BHs undergo kink instability and disrupt the ordered angular momentum supply towards the BH. This can cause the disk misalignment. Here, the snapshots represent the times when the BH-disk system leaves the MAD state (i.e., $\phi$ drops well below $\sim 50$ in Fig.~\ref{fig:plt_vs_time}b) either temporarily (in p1 model) or permanently (in p1N0 model).}
    \label{fig:rho_xz_tilt}
\end{figure*}

Slowly spinning BHs produce jets with much lower luminosities than rapidly spinning BHs. This leads to differences in jet propagation into the infalling stellar material. As the jets drill through a medium with a shallow density profile, (characterized with a lower power-law index, $\rho\propto r^{-\alpha_{\rm p}}$ with $\alpha_{\rm p} < 2$), they are prone to becoming progressively more unstable with distance and eventually falling apart due to the global magnetic kink instability, which causes the jets to twist onto themselves and fall apart~\citep{Bromberg_Tchekhovskoy_2016MNRAS.456.1739B, lalakos_jets_2024}. All else equal, the weaker the jet, the more unstable it is. Fig.~\ref{fig:plt_vs_time} shows that the p\# models, whose progenitors feature such a shallow density slope, $\alpha_{\rm p} = 1.5$, exhibit higher variability in the integrated quantities near the BH, $\mdotRM, \phi, L_{\rm EM}$. This, consequently, results in a highly variable spin-up parameter $s$, especially at low BH spins, $a \lesssim 0.2$, when the jets are at their weakest. For instance, as Fig.\ref{fig:plt_vs_time}(b) shows, in the models with $a=0.1$, e.g., models p1 and p1N0, dimensionless magnetic flux, $\phi$, occasionally drops below the MAD threshold value. As the system leaves the MAD state, BH spindown becomes less efficient. In contrast, the c\# model progenitors, with steeper density profiles, $\alpha_{\rm p} = 2.5$, do not show such strong variability and do not exit the MAD state. Fig.~\ref{fig:rho_xz_tilt} illustrates that the accretion disks can develop significant misalignment with respect to the BH midplane. Such misalignment can occur due to the jets undergoing the global kink instability, disrupting the ordered angular momentum supply near the BH, and causing the accretion disk to wobble and leave the MAD state. A similar effect is present, but less pronounced, in MADs around rapidly spinning BHs, possibly due to the stronger BH spin-disk orientation coupling, or ``magneto-spin'' alignment, at high BH spins \citep{Chatterjee_2023arXiv231100432C}.
These less stable jets spread their energy over large solid angles. This loss of focus can lead to longer breakout times, lower observed GRB luminosities, or even failed jets (see \citealt{Gottlieb2022ApJ...wobbles} for a related discussion).

\subsection{Neutrino-cooled MADs spin evolution}
\label{sec:fits}

\begin{figure}[!tbp]
    \centering
    \includegraphics[width=\columnwidth]{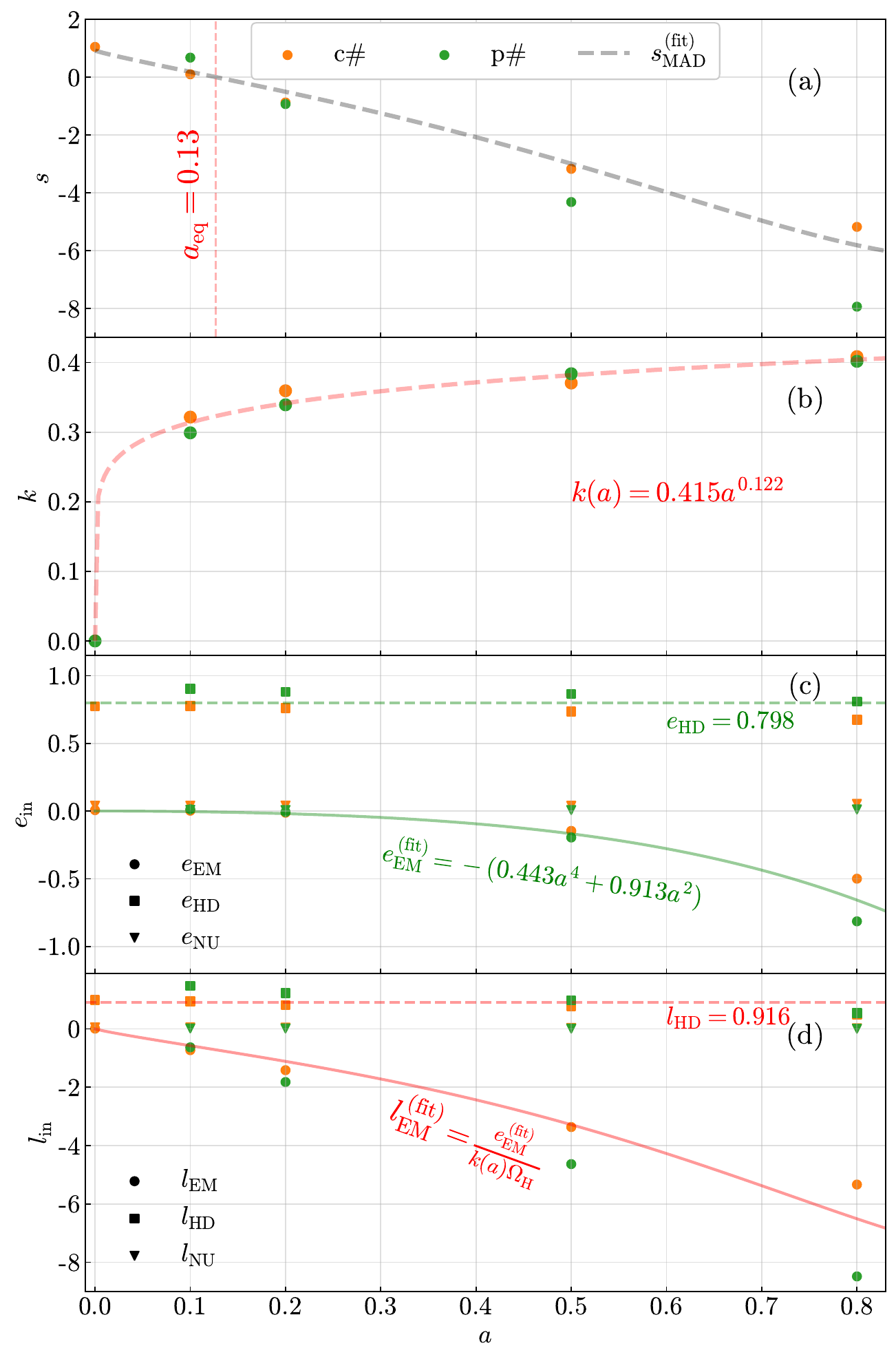}
    \caption{Our model for the spin-up parameter, $s$, well agrees with the simulated data points (panel a) for the model parameters fitted to the simulation (panels b-d). BH spin evolution in the MAD state depends on several key physical values, which we approximate as functions of the spin, $a$. 
    \textbf{Panel (a)} shows the values of the spin-up parameter, $s(a)$ (Eq.~\ref{eq:spinup_parameter}), time averaged over the window, $t > 0.6\,\text{s}$, which is shown in Fig.~\ref{fig:plt_vs_time}(f). Progenitors with the constant density core (c\# models) are marked in orange, and models with the power-law density profile (p\# models) in green. By plugging the fits for values shown in panels (b,c,d) into Eq.(\ref{eq:spinup_parameter}), we compute the model prediction for the MAD spin-up parameter, $s_{\rm MAD}^\text{(fit)}(a)$, shown by the gray dashed line. The equilibrium spin, $\aeq = \aeqval$, which occurs at $s(a) = 0$, is indicated by the vertical red dashed line.
    \textbf{Panel (b)} shows the ratio of the jet-to-event horizon angular frequencies, $k = \Omega_{\rm F} / \Omega_{\rm H}$, with the best fit, $k(a) = 0.415\, a^{0.122}$ (red dashed line).
    \textbf{Panels (c)} and \textbf{(d)} show specific energy, $\ein$, and angular momentum fluxes, $\lin$, which we break into \textit{gas} (HD, marked with squares), \textit{electromagnetic} (EM, circles) and \textit{neutrino} (NU, triangles) contributions. HD components of $\ein,\,\lin$ are approximately independent of the BH spin, $e_{\rm HD}=0.80$, $l_{\rm HD}=0.92$. The EM component of $\ein$, $e_{\rm EM}$, is fitted with an even, 4th degree polynomial (shown with solid green line in panel c). The fit for $l_{\rm EM}$ combines fits for $k(a)$ and $e_{\rm EM} (a)$ (solid red line in panel d). We find that neutrino contributions to $\ein,\,\lin$ are negligible at all spins.}
    \label{fig:el_vs_a}
\end{figure}

In the MAD state, BH spin evolution occurs rapidly due to magnetic field torques on the BH \citep{Lowell2024_spindown}. Here we aim to compute key parameters of the MAD spin evolution model, developed by \cite{Lowell2024_spindown}, and estimate the equilibrium spin, $\aeq$, at which the spin-up parameter vanishes, $s(\aeq)=0$. 
We compute the spin-up parameter in the MAD state, $s_{\rm MAD}$, by time-averaging $s$ over $t > t_{\min}$ (area marked with green on Fig.~\ref{fig:plt_vs_time}), and plot it as a function of $a$ in Fig.~\ref{fig:el_vs_a}(a), for each model. Because the MAD onset time, $t_{\rm MAD}$, is $0.3 \text{ or } 0.4\,\rm{s}$ for c\# or p\# models (see Fig.~\ref{fig:mdot_phi_avg}(b)), we choose to take the average of $s$ starting at $t_{\min}=0.6\,\rm{s}$ to ensure that we capture the spin evolution well in the MAD regime. 

Fig.~\ref{fig:el_vs_a}(a) shows the time-average spin-up parameter in the MAD state, $s_{\rm MAD}$, where all c\# models are marked with orange markers, and all p\# models - with green. We note that there is a slight variation in the values of $s_{\rm MAD}$ between different progenitor models, especially at higher spins, $a \gtrsim{0.5}$. The differences arise because p\# models produce more efficient jets that extract more angular momentum. We speculate that this could be linked to the density profile, where a steeper gradient may hinder jet efficiency by altering the jet's transverse equilibrium. However, this difference is less significant in the vicinity of $s\approx 0$, in the region of our interest for estimating $\aeq$.

We compute the fit of $s_{\rm MAD}(a)$ from Eq.(\ref{eq:spinup_parameter}) using the fits in Fig.\ref{fig:el_vs_a}(b)-(d), shown with gray dashed line in Fig.~\ref{fig:el_vs_a}(a), and find the equilibrium spin, $\aeq = \aeqval$. We find that this value is higher than in thick, non-cooled MADs \citep{narayan_jets_2022, Lowell2024_spindown}, but lower than in radiatively cooled MADs \citep{Lowell_prep}. In the rest of this subsection, we detail the computation of the MAD spin-up parameter, including the key model parameters.

Following \cite{Moderski1996MNRAS..spinAGN, Lowell2024_spindown, Lowell_prep}, we decompose the spin-up parameter, $s$, into contributions due to gas (HD), electromagnetic (EM), and neutrino (NU) components:
\begin{equation}
    s = l_{\rm HD} + l_{\rm NU} - 2a \left( e_{\rm HD} + e_{\rm NU} \right) + l_{\rm EM} - 2a e_{\rm EM} .
    \label{eq:s_decomposed}
\end{equation} 
In the MAD state, the EM components of the specific angular momentum and energy fluxes, $l_{\rm EM}$ and $e_{\rm EM}$, can be parametrized by,
\begin{equation}
    l_{\rm EM} \approx -\frac{\eta_{\rm EM}}{\Omega_{\rm F}}, \quad e_{\rm EM} = -\eta_{\rm EM} ,
    \label{eq:eEM_lEM}
\end{equation}
where $\eta_{\rm EM}$ is the jet-like outflows launching efficiency in Eq.(\ref{eq:eta_jet}), and $\Omega_{\rm F}$ is the Ferraro rotational frequency of the field lines. We can then recast the equation Eq.(\ref{eq:s_decomposed}) and get $s_{\rm MAD}$,
\begin{equation}
    s_{\rm MAD} = l_{\rm HD} + l_{\rm NU} - 2a \left( e_{\rm HD} + e_{\rm NU} \right) - \eta_{\rm EM} \left( \frac{1}{\Omega_{\rm F}} - 2a\right) .
    \label{eq:sMAD}
\end{equation}
Now, we will compute approximations for each of the terms on the r.h.s. of Eq.(\ref{eq:sMAD}). We compute 
\begin{equation}
    k = \frac{\langle\Omega_{\rm F}\rangle_{t,\theta,\varphi}}{\Omega_{\rm H}} \Bigg|_{r=\rhor} ,
    \label{eq:k_def}
\end{equation}
which is the ratio of the angular frequency of the jet to the angular frequency of the BH event horizon, $\Omega_{\rm H}=a/(2\rhor)$, and averaged in $\theta, \varphi$, and $t$, over the same time range as for calculating the average $s_{\rm MAD}$. Fig.~\ref{fig:el_vs_a}(b) shows that the values of $k(a)$ are similar for the progenitor structures and can be well-approximated with a power-law fit (red dashed line),
\begin{equation}
    k^{\rm (fit)}(a) = 0.416\, a^{0.122}.
    \label{eq:k_vs_a_fit}
\end{equation}
Fig.~\ref{fig:el_vs_a}(c),(d) shows different components of specific fluxes that make up $\ein$ (panel [c]) and $\lin$ (panel [d]). Similar to \cite{Lowell2024_spindown}, we approximate the gas (HD) contributions to be independent of BH spin and show them with the dashed lines in Fig.~\ref{fig:el_vs_a}(c),(d), 
\begin{equation}
    e_{\rm HD}^{\rm (fit)} \approx 0.798, \quad l_{\rm HD}^{\rm (fit)} \approx 0.916 .
    \label{eq:eHD_lHD_fit}
\end{equation}
The EM contribution to $\ein$, negative-signed $\eta_{\rm EM}$, is well-fitted by an even quartic function (solid green line in Fig.~\ref{fig:el_vs_a}c),
\begin{equation}
    -e_{\rm EM}^{\rm (fit)} = \eta_{\rm EM}^{\rm (fit)} = 0.443 a^4 + 0.913 a^2 .
    \label{eq:eta_jet_fit}
\end{equation}
Once we compute the fits of $k(a)$ and $\eta_{\rm EM}(a)$, we can calculate the fit of the EM contribution to specific angular momentum, $l_{\rm EM}$ (Eq.~\ref{eq:eEM_lEM}), shown by the solid red line in Fig.~\ref{fig:el_vs_a}(d). Fig.~\ref{fig:plt_vs_time}(c),(d) shows that the jet and neutrino efficiencies, $\eta_{\rm EM}$ and $\eta_{\nu}$, are comparable at lower spins, $a \lesssim 0.2$. Nevertheless, we find that the neutrino contribution (NU, marked by triangles in Fig.~\ref{fig:el_vs_a}(c),(d)) to the specific fluxes is very small, and we neglect it in our analysis. Now, together with the fits, $k^{\rm (fit)}(a)$ in Eq.(\ref{eq:k_vs_a_fit}) and $\eta_{\rm EM}^{\rm (fit)}$ in Eq.(\ref{eq:eta_jet_fit}), we can compute $s_{\rm MAD}$ from Eq.(\ref{eq:sMAD}),
\begin{equation}
    s_{\rm MAD}^{\rm (fit)} (a) = l_{\rm HD}^{\rm (fit)} - 2a e^{\rm (fit)}_{\rm HD} - \eta^{\rm (fit)}_{\rm EM} (a) \left[ \frac{2\rhor(a)}{a} \frac{1}{k^{\rm (fit)}(a)} - 2a \right] .
    \label{eq:sMAD_fit}
\end{equation}
The gray dashed line in Fig.~\ref{fig:el_vs_a}(a) shows how this fit provides a reasonable approximation to $s_{\rm MAD}(a)$ measured in the simulations.

\subsection{How neutrinos affect BH spin evolution}\label{sec:neutrinosandspin}

Fig.\ref{fig:plt_vs_time} shows that models without neutrino transport demonstrate some noticeable differences in dynamics near the BH. Model p1N0 (shown with thick red transparent curves) has a similar mass accretion rate profile as the neutrino-cooled models with the same initial progenitor structure (Fig.\ref{fig:plt_vs_time}a), however, it does not stay in the MAD state (Fig.\ref{fig:plt_vs_time}b), leading to lower jet power (Fig.\ref{fig:plt_vs_time}d). Model c1N0 (shown with thick red dashed transparent curves) has a much shallower mass accretion rate temporal profile than other models with the same constant-density progenitor structure (Fig.\ref{fig:plt_vs_time}a), which results in the MAD onset time that is significantly longer ($\sim 1$~s vs.{} $0.3$~s). Fig.\ref{fig:plt_vs_time}(f) shows that the values of the spin-up parameter, $s$, are comparable between p1 and p1N0 models (with and without the neutrino transport at spin $a=0.1$). However, because p1N0 model never enters the MAD state, its spin-up parameter is not representative of the MAD state. In model c1N0, $s$ is lower and negative after the MAD onset, compared to positive $s$ in model c1: this illustrates the importance of neutrino cooling, which increases the BH spin equilibrium value.

\section{Spin evolution analysis}
\label{sec:spin_evol_analysis}

\begin{figure*}[!tbp]
    \centering
    \includegraphics[width=\textwidth]{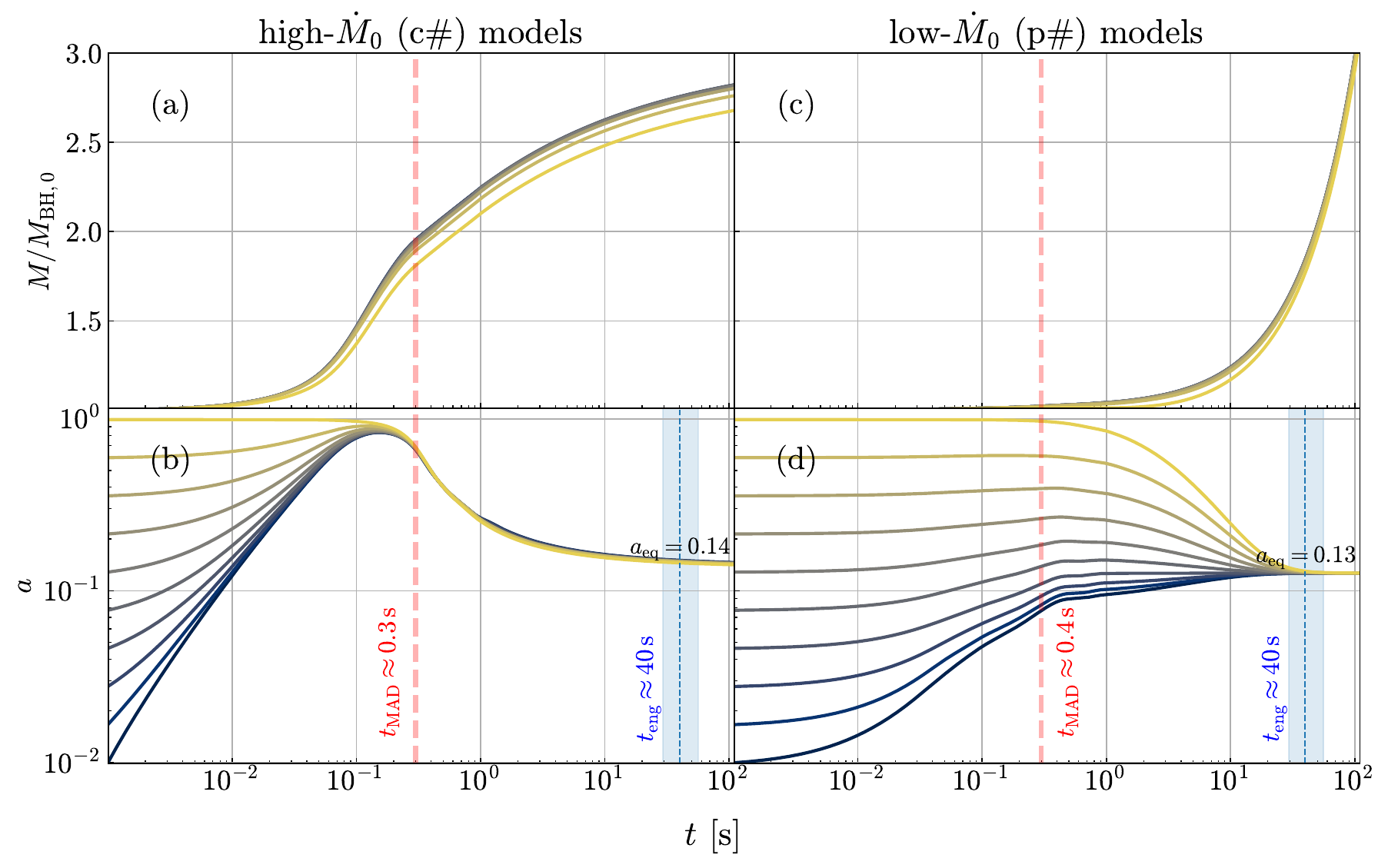}
    \caption{We model the time evolution of the BH mass (top row) and spin (bottom row) by integrating Eqs.~\ref{eq:spin_evol_ODEs} on the timescale of the BH engine activity, $t_{\rm eng}$ (Eq.~\ref{eq:t_eng}) that are well beyond our simulation runtimes, $t_{\rm sim} \sim 1\,\rm{s}$. 
    Left column shows the solution, $M(t)$ and $a(t)$, for high-$\mdotRM$, constant density core models. Right column shows the solution for more typical, low-$\mdotRM$ collapsar progenitors, for which we use the fits of $\mdotRM$, $\phi$ shown in Fig.~\ref{fig:mdot_phi_avg}. 
    High-$\mdotRM$ progenitors accrete rapidly before reaching MAD state and reach high spin values at MAD onset, $t_{\rm MAD} \approx 0.3\,\rm{s}$. However, in the MAD state, the powerful magnetically-driven outflows efficiently spin-down the BH to $a \approx 0.14$ before the gas supply is depleted, independent of initial spin, $0<a_0<1$. 
    Low-$\mdotRM$ progenitors initially accrete slower, but acquire much more mass in the MAD state, at $t>t_{\rm MAD} \approx 0.4\,\rm{s}$. 
    This results in a slower spin evolution initially; however, on the timescale of BH engine activity, $t_{\rm eng} \gtrsim{} 40$~s, the spin eventually approaches its equilibrium value, $\aeq = 0.13$, consistent with the value obtained in Sec.~\ref{sec:spin_evol}.}
    \label{fig:ODE_solution}
\end{figure*}

Once the MAD state is reached, it will likely continue launching powerful EM outflows. In the progenitor models that we consider in this study, long-term mass accretion rate $\mdotRM$ goes as $t^{-p}$, where $p = 0.06$ for low-$\mdotRM$ (p\#) models and $p=1.30$ for high-$\mdotRM$ (c\#) models (see Fig.~\ref{fig:mdot_phi_avg}). As $\mdotRM$ decreases, magnetic flux continuously escapes off the BH such that the remaining flux keeps the BH in the MAD state, $\phi \sim \phi_{\rm MAD}$. As long as the BH does not accrete any magnetic flux of opposite polarity, we assume that it stays in MAD state until the magnetic flux supply depletes. Assuming that these successful collapsar explosions give rise to GRBs, and the GRB duration time represents the BH engine activity timescale, we can make predictions of the final BH spins.

Total BH spin evolution timescale, which should be the BH engine activity timescale, $t_{\rm eng}$, is the sum of the time it takes for the jet to break out of the stellar mantle, $t_{\rm br}$, and the observed duration of the resulting GRB, $t_{\rm GRB}$
\begin{equation}
    t_{\rm eng} = t_{\rm br} + t_{\rm GRB} .
    \label{eq:t_eng}
\end{equation}

We approximate the breakout time to be $t_{\rm br} \lesssim 10\, \rm{s}$ \citep[see e.g.,][]{Gottlieb2023ApJ...natalspin}, and $\log_{10}t_{\rm LGRB} = 1.476 \pm 0.189$, corresponding to $t_{\rm LGRB, mean} \approx 30\,\rm{s}$ \citep[][the latest Fermi-GBM catalog]{FermiGBM_GRBcatalog_von_Kienlin_2020}. Hence, we adopt $t_{\rm eng} \lesssim 40\,\rm{s}$.

Our simulations are much shorter than $t_{\rm eng}$. Therefore, we model the BH spin evolution using the approach in \cite{Jacquemin-Ide2024ApJ...collapsarspindown} \citep[also see][]{Wu_2024_GRB_spinevol}, where they solve a system of spin evolution ODEs using the fits of the jet launching efficiency, $\eta_{\rm EM} (a)$, and jet-to-BH event horizon angular frequency ratio, $k(a)$, derived in \cite{Lowell2024_spindown}, which used simulations of thick MADs $(h/r\sim 0.3)$. In the context of collapsars, where MADs are neutrino-cooled, we use the fits derived in Sec.~\ref{sec:results}, namely Eqs.~\ref{eq:k_vs_a_fit}, \ref{eq:eta_jet_fit}.

\cite{Jacquemin-Ide2024ApJ...collapsarspindown} bridge the BH spin evolution of a weakly magnetized disk (before MAD onset) and MAD. They combine the BH spin evolution equations of the standard, thin, Novikov-Thorne (NT) disk model \citep{novikov_astrophysics_1973}, whose spin evolves according to \cite{Bardeen_1970Natur.226...64B}, and spin evolution in MAD regime, from \cite{Lowell2024_spindown},
\begin{equation}
\begin{split}
    \frac{da}{dt} &= \frac{\mdotRM(t)}{M} \Big[ s_{\rm NT} (1 - w_{\rm MAD}(t)\,) + s_{\rm MAD} \,w_{\rm MAD}(t)\, \Big] , \\
    \frac{dM}{dt} &= \mdotRM(t) \Big[ e_{\rm NT} \left( 1 - w_{\rm MAD}(t)\, \right) + \left( e_{\rm HD} - \eta_{\rm EM}\right) w_{\rm MAD}(t) \Big] ,
    \label{eq:spin_evol_ODEs}
\end{split}
\end{equation}
where the expressions for $e_{\rm NT}, l_{\rm NT}$ are given by Eq.~\ref{eq:NT_ein_lin}, and $s_{\rm MAD}(a)$ is given by Eq.~\ref{eq:sMAD}.

We introduce the weight $w_{\rm MAD}(t)$ to account for spin evolution in both pre-MAD and MAD regimes. $w_{\rm MAD}(t)$ represents the relative 'weight' of the MAD state, and ranges from $0$ when it is far from MAD ($\phi \ll \phi_{\rm MAD}$) to $1$ when it is fully MAD ($\phi = \phi_{\rm MAD}$). We set $w_{\rm MAD} (t) = \min [1, \phi(t) / \phi_{\rm MAD}]^2$, given that the BZ jet power is proportional to square of the BH magnetic flux, $L_{\rm BZ} \propto \Phi^2$ \citep{BZ_1977}. We use the progenitor-averaged profiles of $\mdotRM(t)$ and $\phi(t)$ computed directly from the simulations (see Fig.~\ref{fig:mdot_phi_avg}).

We solve Eqs.~\ref{eq:spin_evol_ODEs} for a range of initial spins $a_{0} \in (10^{-2}, 1)$, which we show in Fig.~\ref{fig:ODE_solution}. Fig.~\ref{fig:ODE_solution}(a),(b) shows that in high mass accretion rate progenitors (c\# models), before the MAD onset (marked by a vertical dashed red line), the BH spins up as it rapidly accretes in the absence of BZ jet. After $t_{\rm MAD} \sim 0.3\rm{s}$, spin evolution is similar for all initial spins. For a typical engine activity duration $t_{\rm eng} > 10\, \rm{s}$, the final spin is $a_{f} = 0.14$, consistent with $\aeq$. Fig.~\ref{fig:ODE_solution}(c),(d) shows that in typical collapsar progenitors (p\# models), mass accretion rates are lower, therefore BH spin converges towards its equilibrium value much more slowly. Here, the final spin will be determined based on the initial spin $a_0$ and engine duration time $t_{\rm eng}$. Spin evolution happens faster if the initial spin is lower (i.e. closer to the equilibrium spin $\aeq$). For engine duration $t_{\rm eng} > 40\,\rm{s}$, final spin lies in range $a_f \in (0.1-0.2)$; consistent with $\aeq \approx \aeqval$ obtained in Sec.~\ref{sec:results}. 

In summary, for a wide range of collapsar progenitor models with mass accretion rates, $\mdotRM \sim 0.1 - 10\,\msun/\text{s}$, neutrino-cooled disks and magnetically-driven outflows, we expect the final spin to be universally around the equilibrium spin $\aeq \sim \aeqval$.

\section{Discussion and Conclusion}
\label{sec:discussion}

\subsection{Summary}

In this study, we performed global 3D GRMHD simulations of collapsars with neutrino-cooled disks to investigate collapsar BH spin evolution. Collapsar BHs are strong candidates for LGRBs and potential progenitors of GW sources. BH spin evolution is crucial, as it can directly impact the jet duty cycle, and affect the energetics of the jets. Moreover, observations of the GRBs point at the existence of an upper energy cutoff \citep{atteia_grb_2025}. Given that the BH spin sets the maximum power achievable by the jets, this cutoff may be governed by BH spin dynamics. 

To study the spin evolution across a range of physical conditions (e.g., mass accretion rates), we simulate two types of collapsar progenitor models with different stellar density profiles: c\# models, which feature a constant-density core and have been shown to favor $r$-process element production, and p\# models, which follow a typical collapsar profile with \( \rho (r) \propto r^{-1.5} \), but are less efficient at generating $r$-process elements \citep{Issa2024arXiv241002852I...nuHAMR}.

Consistent with previous work, we find that a near-MAD state is required for jet-like outflow launching, with a dimensionless flux of \(\phi \gtrsim{} 10\) \citep{Komissarov_2009MNRAS.397.1153K,gottlieb_black_2022,Issa2024arXiv241002852I...nuHAMR}. Once launched, the disk-jet system rapidly reaches the MAD state on a timescale of \( t_{\rm MAD} \sim 0.3-0.4 \) s, after which all angular momentum and energy fluxes stabilize around the BH, such that spin-up parameter $s$ reaches the steady-state values (see Fig.~\ref{fig:plt_vs_time}). At this stage, we measure these fluxes (magnetic, hydrodynamic, and neutrino) to determine the torques exerted by the jets and disk on the BH, represented by the spin-down parameter, \( s \) (see Fig.~\ref{fig:el_vs_a}). In agreement with \cite{Lowell2024_spindown}, we find that BH spin-down is primarily driven by electromagnetic jet torques, while hydrodynamic disk torques provide a smaller yet non-negligible contribution. We find that neutrino fluxes at the BH are entirely negligible, as they are too weak to influence spin evolution compared to other torques. However, as discussed below and in Sec.~\ref{sec:neutrinosandspin}, neutrinos remain important because their cooling effects modify the strength of both electromagnetic and hydrodynamic torques.

For the first time in collapsar simulations with neutrino transport, we find that MADs efficiently spin down their BHs. Under the influence of MAD torques, the BH reaches a low equilibrium spin of \( \aeq = \aeqval \). Following \cite{Lowell2024_spindown}, we construct a semi-analytical spin-down model, which accurately reproduces the fluxes and torques governing BH spin evolution in collapsars.

Surprisingly, we find that the evolution of the accretion rate (\( \mdotRM \)) and magnetic flux (\( \phi \)) at the BH surface remains largely independent of the initial spin. In contrast, jet efficiency varies by more than an order of magnitude. This suggests that the magnetic flux accumulation (and the formation of the MAD state), rather than the jet power (which is set by $a$), primarily governs gas dynamics near the BH. However, at later times (\( t \geq 1.25\, \rm{s} \)), we observe long-period oscillations in the dimensionless magnetic flux, \( \phi \), and jet efficiency, \( \eta_{\rm EM} \), for very low spins, \( a = 0.1 \).

The general behavior of \( \mdotRM(t) \) and \( \phi(t) \) allows us to develop a time-dependent spin evolution model. Following \cite{Jacquemin-Ide2024ApJ...collapsarspindown}, we compute the BH spin evolution, \( \dot{a} \), and find that both models (c\# and p\#) converge to a final spin of \( a \sim \aeqval \) after a typical LGRB engine lifetime, \( t_{\rm eng} \sim 40 \) s \citep[][see also Section \ref{sec:spin_evol_analysis}]{FermiGBM_GRBcatalog_von_Kienlin_2020}. In both cases, the system has sufficient time to accrete enough mass to nearly reach equilibrium spin. Overall, this work improves upon \cite{Jacquemin-Ide2024ApJ...collapsarspindown} by incorporating a more self-consistent spin evolution model, accounting for neutrino cooling and more realistic dependencies of magnetic flux and accretion rate evolution, ultimately providing better constraints on BH spin evolution and final spin.

\subsection{Comparison with other work}

We find a $2-4$ times larger value of the equilibrium spin, \( \aeq = \aeqval \), than in the previous spin-down models of adiabatic non-radiative MADs, \( \aeq = 0.035 \) \citep{narayan_jets_2022} and \( \aeq = 0.07 \) \citep{Lowell2024_spindown}. This discrepancy may stem from the lower disk thickness in our simulations due to neutrino cooling. 
Notably, \citet[][in preparation]{Lowell_prep}  find that equilibrium spin increases as disk thickness decreases. Our result of \( \aeq = \aeqval \) falls between their predictions of \( \aeq = 0.07 \) for adiabatic, thick MADs and \( \aeq \approx 0.3 \) for radiativelly-cooled, thin MADs ($h/r = 0.1$) and further reinforces their trends.

\cite{ricarte_recipes_2023} measured BH spin-down in relatively thick (\( h/r \geq 0.18 \)) radiative super-Eddington MADs. Although cooling processes differ between our models and theirs, we can compare cases with roughly similar disk thicknesses. For \( h/r \simeq 0.25 \), they find \( \aeq \sim 0.18 \), which aligns with our results. However, for \( h/r \simeq 0.2 \), they report \( \aeq \sim 0.8 \), a large discrepancy from our findings.

\cite{Wu_2024_GRB_spinevol} argue that MADs are unlikely to produce LGRBs, as their narrow progenitor mass distribution would limit the observed diversity in LGRB energies. Using the models of \cite{Lowell2024_spindown} and \cite{Jacquemin-Ide2024ApJ...collapsarspindown}, they suggest that most (subluminous) LGRBs can instead be powered by systems with much lower magnetic flux saturation levels, $\phi \ll \phi_{\rm MAD}$. However, this contrasts with simulations showing that jet launching requires dynamically important magnetic fields in a near-MAD state. In our framework, slowly spinning BHs produce unstable jets, which may reduce energy deposition into the stellar envelope and lead to the lower jet energies observed at breakout, potentially explaining subluminous LGRBs (see also below and in Sec.~\ref{sec:wobbling}).
\cite{Wu_2024_GRB_spinevol} further propose that a sub-MAD state ($\phi \sim 10$) could produce the most luminous GRBs, as the higher equilibrium spin ($\aeq \sim 0.5$) would enhance jet power. This is an interesting possibility, that is in the same direction as our work finding a higher equilibrium spin for neutrino cooled collapsars. 

A striking result that we uncover is that for low spin, models p1 and c1 at $a = 0.1$, we observe long-term oscillations in the dimensionless BH magnetic flux, \( \phi \), and jet efficiency, \( \eta_{\rm EM} \), at late times (\( t \geq 1.25\, \rm{s} \)). Interestingly, the p1 model shows stronger variability than the c1 model and during the times when the jets are exceptionally weak. The main difference between these two models is the p1 model's shallower density profile, which makes the jets progressively less stable as they propagate outwards \citep{Bromberg_Tchekhovskoy_2016MNRAS.456.1739B}. Also consistent with this picture is that weaker jets are less stable.
It is possible that this instability perturbs the large scale gas flow, which in turn perturbs the accretion disk and affects the jet launching.
Indeed, the variability appears to be linked to variations in the disk's angular momentum vector, similar to the wobbles reported by \citet{Gottlieb2022ApJ...wobbles} and \citet{lalakos_jets_2024}. The misalignment between the disk and BH angular momentum vectors induces jet wobbles. While \cite{Gottlieb2022ApJ...wobbles} also report jet wobbles at the BH photosphere, it remains unclear whether these oscillations affect electromagnetic efficiency at the event horizon in their study. Moreover, such variations in \( \eta_{\rm EM} \) are absent in prior studies of weakly spinning MAD collapsars \citep{Gottlieb2023ApJ...natalspin}. 
Quantifying the variability in \( \eta_{\rm EM} \), \( \phi \) and $L_{\rm EM}$ at both the event horizon and the photosphere is crucial, as they may have significant observational implications. 

\subsubsection{Observational implications}

BH spin-down for initially rapidly-spinning BHs results in the strongest jet power occurring while the jets are still inside the stellar envelope. As the jet emission is observable only post-breakout, understanding spin evolution and its timing relative to the breakout time (\( t_{\rm b} \)) is crucial for interpreting GRB observations. However, our simulations do not extend to the jet breakout and lack self-consistent spin evolution modeling, which limits our ability to fully capture these effects. Nonetheless, we find that BH spins would settle below \( a \leq 0.2 \) in most models by \( t \sim 5 \) s, comparable to \( t_{\rm b} \). These slowly-spinning BHs produce jets with moderate energies (\( L_{\rm j} \leq 10^{52} \) erg/s), more in line with typical observed GRB powers, although, at the higher end of the distribution \citep{goldstein_estimating_2016}. Note that in this study we consider a rather massive $70 M_\odot$ progenitor: for a more typical, five times smaller, $14M_\odot$ progenitor star, the jet power would be $5$ times smaller and the spin-down time $5$ times longer. The suppression of higher jet power due to BH spin-down before \( t_{\rm b} \) may explain the observed upper limit on GRB luminosities \citep{Jacquemin-Ide2024ApJ...collapsarspindown,Wu_2024_GRB_spinevol,atteia_grb_2025}. Notably, for an initial spin of \( a = 0 \), weaker jets are more easily achieved, as the BH is not spun down, but is spun up instead \citep{Jacquemin-Ide2024ApJ...collapsarspindown}, resulting in the jet powers that are more consistent with the observed GRB energy distribution.

GW observations of BBH mergers provide increasingly precise insights into BH spin. Multiple Bayesian population studies indicate that the spin distribution is centered around low values, \( a \sim 0.1-0.4 \), with high spins (\( a \geq 0.6 \)) being rare\footnote{see however, \cite{galaudage_building_2021}} or absent \citep{the_ligo_scientific_collaboration_gwtc-3_2021,galaudage_building_2021,the_ligo_scientific_collaboration_population_2022,callister_no_2022,tong_population_2022,edelman_cover_2023}. If LGRBs are progenitors of BBH mergers, our models naturally explain the observed spin distribution in GW events.

Whether collapsars are progenitors of BBH mergers remains an open question. Population studies comparing BBH and LGRB redshift distributions, and their event rates, suggest that they are distinct populations due to their different occurrence frequencies \citep{arcier_are_2022}. However, \cite{bavera_probing_2022} used binary evolution models to show that tidal interactions in close binaries can explain a subpopulation of spinning, merging BBHs associated with LGRBs, predicting $\sim 10$\% of GWTC-2 events. This aligns with Bayesian analyses of delay times between LGRBs and BBH mergers, which find that up to $14$\% of LGRBs could be BBH progenitors, provided the BHs are slowly spinning; while for rapidly-spinning BHs, this fraction drops to at most $1.3$\% \citep{wu_are_2024}.

\subsection{Future work}

The main limitations of our work are the short evolution timescales and the lack of self-consistent spin evolution modeling. With the  simulation runtimes under \(3\) s, we cannot constrain the breakout time, \(t_{\rm b}\), in our models. Determining \(t_{\rm b}(a)\) would be highly valuable, as it would help refine jet power estimates by providing \(L_{\rm EM}(t_{\rm b})\), and offer insights into whether observed jet power evolves significantly — potentially conflicting with GRB variability constraints \citep{mcbreen_cumulative_2002}. A more consistent approach would involve dynamically evolving the BH spin within the GRMHD simulation, allowing us to correctly account for how changes in \(a\) and \(\eta_{\rm EM}\) impact \(t_{\rm b}\). This would not only improve modeling of jet power pre- and post-breakout, but also clarify whether spin reaches equilibrium before \(t_{\rm b}\). The effects of spin evolution on collapsar dynamics and jet power evolution remain largely unexplored \citep[see however, 2D simulations by][]{shibata_outflow_2024}.  In this work, we studied a rather massive progenitor star of $70M_\odot$.  Lower mass stars feature lower mass accretion rate and longer spin evolution timescales. We leave the study of such stars to future work.

In conclusion, this work advances collapsar modeling by providing a more self-consistent approach to BH spin evolution, by incorporating neutrino transport and realistic accretion dynamics. These results have important implications for understanding the power and variability of GRBs, as well as the spin evolution of BHs in GW progenitor systems.

\begin{acknowledgements}
\section*{Acknowledgements}
DI thanks Ore Gottlieb, Aretaios Lalakos for useful comments and discussions.
DI is supported by Future Investigators in NASA Earth and Space Science and Technology (FINESST) award No. 80NSSC21K1851. 
BL acknowledges support by a National Science Foundation Graduate Research Fellowship under Grant No. DGE-2234667. BL also acknowledges support by a Illinois Space Grant Consortium (ISGC) Graduate Fellowship supported by a National Aeronautics and Space Administration (NASA) grant awarded to the ISGC. JJ acknowledges support by the NSF AST-2009884, NASA 80NSSC21K1746 and NASA XMM-Newton  80NSSC22K0799 grants. AT acknowledges support by NASA 
80NSSC22K0031, 
80NSSC22K0799, 
80NSSC18K0565 
and 80NSSC21K1746 
grants, and by the NSF grants 
AST-2009884, 
AST-2107839, 
AST-1815304, 
AST-1911080, 
AST-2206471, 
AST-2407475, 
OAC-2031997. 
This work was performed in part at the Kavli Institute for Theoretical Physics (KITP) supported by grant NSF PHY-2309135.
This work was performed in part at Aspen Center for Physics, which is supported by National Science Foundation grant PHY-2210452.
This research used resources of the National Energy Research Scientific Computing Center, a DOE Office of Science User Facility supported by the Office of Science of the U.S. Department of Energy under Contract No. DE-AC02-05CH11231 using NERSC allocations m4603 (award NP-ERCAP0029085) and m2401. The computations in this work were, in part, run at facilities supported by the Scientific Computing Core at the Flatiron Institute, a division of the Simons Foundation. An award of computer time was provided by the ASCR Leadership Computing Challenge (ALCC), Innovative and Novel Computational Impact on Theory and Experiment (INCITE), and OLCF Director’s Discretionary Allocation programs under award PHY129.  This research was partially carried out using resources from Calcul Quebec (http://www.calculquebec.ca) and Compute Canada (http://www.computecanada.ca) under RAPI xsp-772-ab (PI: Daryl Haggard). This research also used HPC and visualization resources provided by
the Texas Advanced Computing Center (TACC) at The University
of Texas at Austin, which contributed to our results via the LRAC allocation AST20011 (http://www.tacc.
utexas.edu).

\end{acknowledgements}

\bibliography{biblio.bib}

\appendix
\label{appendix}

\section{NT disk model expressions for $\ein$ and $\lin$}

In the Novikov-Thorne (NT) disk model \citep{novikov_astrophysics_1973}, the specific energy and angular momentum fluxes on the BH are given by
\begin{equation} \label{eq:NT_ein_lin}
    \ein = \sqrt{ 1 - \frac{2}{3R_{\rm ms}} }, \quad \lin = \frac{2 M}{3 \sqrt{3}} \left( 1 + 2 \sqrt{3 R_{\rm ms} - 2} \right) ,
\end{equation} 
where $R_{\rm ms}$ is the radius of the marginally stable orbit (and in this case the ISCO),
\begin{equation}
\begin{split}
    R_{\rm ms} &= 3 + Z_2 - \sqrt{(3 - Z_1) (3 + Z_1 + 2 Z_2)} \\
    Z_1 &= 1 + \sqrt[3]{1 - a^2} \left( \sqrt[3]{1 + a^2} + \sqrt[3]{1 - a^2} \right) \\
    Z_2 &= \sqrt{3 a^2 + Z_1^2} .
\end{split}
\end{equation}

\section{Measuring fluxes in the presence of the numerical floors}

\begin{figure}[hbtp]
    \centering
    \includegraphics[width=0.4\columnwidth]{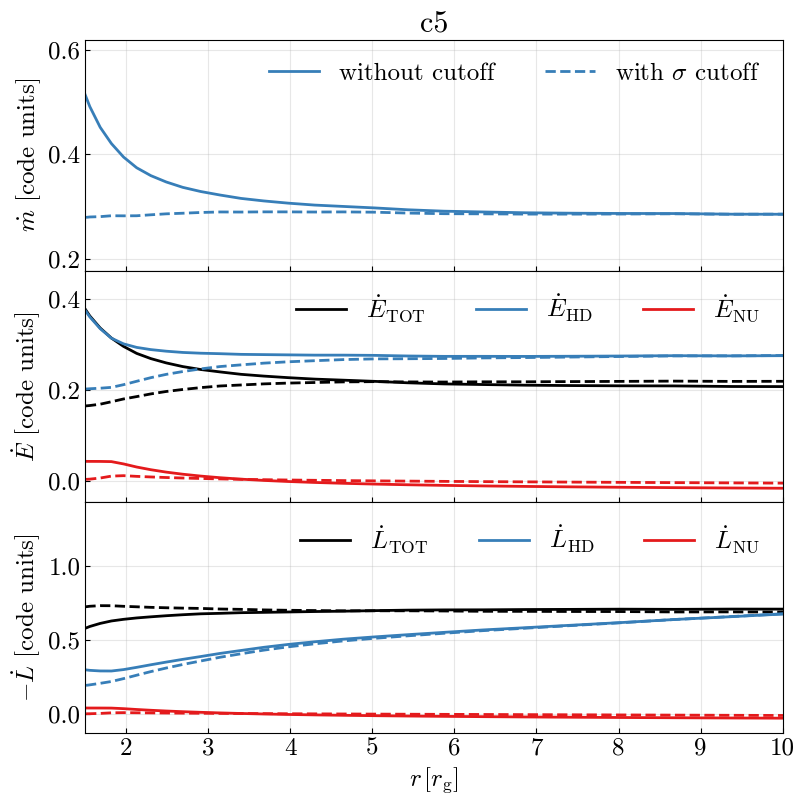}
    \caption{Rest mass (panel [a]), specific energy (panel [b]) and angular momentum (panel [c]) fluxes in model c5, which need to be corrected to minimize the contribution of the numerical density floors. Soild lines show unmodified radial flux profiles and dashed lines show profiles where a magnetization cutoff was applied, $\sigma < (2/3)\sigma_{\max}$.}
    \label{fig:floors_sigmacutoff}
\end{figure}

To minimize the contribution of the numerical density and energy floors to the rest-mass fluxes, hydrodynamical contributions to the specific angular momentum and energy fluxes, we integrate them by only considering regions where $\sigma < (2/3)\sigma_{\max}$, similar to the approach outlined in \cite{Lowell2024_spindown}. As a demonstration, Fig.\ref{fig:floors_sigmacutoff} shows the radial profiles of the fluxes in model c5, averaged over a time window $t\in (2-4)\times 10^4\,\rg/c$, where solid lines show integrated fluxes without any modifications, and dashed lines - fluxes without the contribution of those highly magnetized regions. We see that without the magnetization cutoff, the hydrodynamical fluxes measured near the event horizon, $r=\rhor$ are overestimated due to unphysical addition of density and internal energy. We use modified $e_{\rm HD}$, $l_{\rm HD}$, $e_{\rm NU}$, $l_{\rm NU}$ and $\mdotRM$ measured at the event horizon, and compute $l_{\rm EM}$ and $e_{\rm EM}$ without modifications, as they do not contain neither density nor internal energy terms.

\end{document}